\documentclass[fleqn,10pt]{wlscirep}
\DeclareUnicodeCharacter{00A0}{ }
\DeclareUnicodeCharacter{00A0}{~}
\usepackage{amsmath}
\usepackage{float}
\title{Characterization of Thin Film Materials using SCAN meta-GGA, an Accurate Nonempirical Density Functional}

\author[1,*]{I. G. Buda}
\author[1]{C. Lane}
\author[1]{B. Barbiellini}
\author[2]{A. Ruzsinszky}
\author[3]{J. Sun}
\author[1]{A. Bansil}
\affil[1]{Northeastern University, Physics, Boston, MA 02115, USA}
\affil[2]{Temple University, Physics, Philadelphia, PA 19122, USA}
\affil[3]{The University of Texas at El Paso, Physics, El Paso, TX 79958, USA}
\affil[*]{i.buda@neu.edu}

\begin{abstract}
We discuss self-consistently obtained ground-state electronic properties of monolayers of graphene and a number of 'beyond graphene' compounds, including films of transition-metal dichalcogenides (TMDs), using the recently proposed strongly constrained and appropriately normed (SCAN) meta-generalized gradient approximation (meta-GGA) to the density functional theory. The SCAN meta-GGA results are compared with those based on the local density approximation (LDA) as well as the generalized gradient approximation (GGA). As expected, the GGA yields expanded lattices and softened bonds in relation to the LDA, but the SCAN meta-GGA systematically improves the agreement with experiment. Our study suggests the efficacy of the SCAN functional for accurate modeling of electronic structures of layered materials in high-throughput calculations more generally.

\end{abstract}
\begin{document}

\flushbottom
\maketitle

\thispagestyle{empty}

\section*{Introduction}
\

Discovery of graphene, a one-atom-thick crystal of carbon, has spurred an intense interest in the electronic properties of 2D materials more generally \cite{Novoselov26072005, RevModPhys.81.109}.
Recent research has turned to ‘beyond graphene’ materials, which exhibit novel spin and charge transport properties, including the possibility of harboring quantum spin Hall and other topological phases \cite{RevModPhys.88.021004} relevant for next generation electronics applications and as materials platforms for replacing the current Si-based technologies.
For example, unlike the flat structure of graphene, silicene, germanene and stanene, which are Si, Ge and Sn based cousins of graphene, assume a crystal structure that is naturally buckled \cite{doi:10.1021/ar400180e, 1367-2630-16-9-095002, Zhu2015}.
As a result, these materials exhibit spin-split states, which can be controlled via external electric fields \cite{Tsai2013}.
Phosphorene displays remarkable mechanical flexibility and sensitive tuning of electronic properties by mechanical strain \cite{:/content/aip/journal/apl/104/25/10.1063/1.4885215}. Ultra-thin films of transition metal dichalcogenides (TMDs) undergo a transition from an indirect to a direct band gap semiconductor in the monolayer limit, and have become attractive candidates for nanoelectronics \cite{Wang2012}, water-splitting \cite{Yu2015}, photocatalysis \cite{doi:10.1021/jz502646d} and other applications.
\

\
The need for theoretical methods capable of accurate and efficient prediction of structural and electronic properties of atomically thin films and layered materials is clear. In this connection, improvements in density functional theory (DFT) \cite{PhysRev.136.B864} based first-principles computations, which have been the workhorse in the field for over five decades \cite{zangwill2015half, QUA:QUA22829}, have centered around the development of new classes of exchange-correlation functionals.
One of the latest advances in this direction is the SCAN meta-GGA scheme, which has been proposed recently \cite{PhysRevLett.115.036402}. Our purpose in this study is to assess the efficacy of the SCAN meta-GGA for addressing the ground state properties of 2D materials. SCAN meta-GGA has been tested in diversely bonded systems \cite{sun2016accurate}, where it has been shown to capture a wide range of physical structures without being fitted to any specific type of bonding. These SCAN-based existing studies include: MnO$_2$ polymorphs \cite{PhysRevB.93.045132};  Cu-intercalated birnessite \cite{doi:10.1021/acs.langmuir.5b02936}; and, band gaps of semiconductors and insulators \cite{PhysRevB.93.205205}. Here we show that SCAN meta-GGA yields a systematic improvement over the LDA and GGA (at a comparable cost) in modeling ground state properties of 2D materials. For this purpose, we consider the application of SCAN to monolayers of graphene and a number of 'beyond graphene' compounds, including films of transition-metal dichalcogenides (TMDs) as exemplar 2D systems. 

An outline of this article is as follows. The introductory remarks above are followed by an overview of the SCAN functional and its construction. We then describe the relevant computational details, followed by a presentation and discussion of our results, and a summary of our conclusions.   

\section*{Overview of the SCAN Methodology}  
\ 

Within the framework of the DFT, the total energy of the many-body electron system, \(E_{total}[n]\), can be written in terms of the electron density, \(n\), as 
\newcommand\indentdisplays[1]{%
     \everydisplay{\setlength\displayindent{#1}%
     \addtolength\displaywidth{-#1}}}

\indentdisplays{150pt}  

\begin{equation} 
E_{total}[n]=K+E_{ie}+E_{ee}+E_{xc},
\end{equation}

\noindent where \(K\) is the independent-electron kinetic energy, \(E_{ie}\) is the Coulomb energy between the electrons and ions, \(E_{ee}\) describes the classical electron-electron Coulomb interaction, and \(E_{xc}\) the exchange-correlation energy.
Approximation schemes for \(E_{xc}\) can be arranged conceptually on the rungs of the so-called DFT \emph{Jacob's Ladder} \cite{DFTPerdew} in the sense that this ladder leads to the "heaven" of chemical accuracy. Various rungs of this ladder, beginning with the lowest rung, are:  LDA \cite{PhysRev.140.A1133, vosko1980accurate}; GGA \cite{PhysRevLett.77.3865, barbiellini1990effects}; meta-GGA \cite{tao2003climbing, :/content/aip/journal/jcp/137/5/10.1063/1.4742312}; Hybrid functionals \cite{heyd2003hybrid}; and, finally the random phase approximation (RPA) \cite{ren2012random}. Computational demands, along with the accuracy of the schemes increase as we go up the rungs of the ladder.
\

Formally, the $E_{xc}[n]$ term can be cast as a double integral over space,  which involves half of the Coulomb interaction between electrons and the associated exchange-correlation holes \cite{:/content/aip/journal/jcp/137/5/10.1063/1.4742312, PhysRevB.13.4274}, but it is computationally expensive to evaluate.
In the semilocal approximation, this term is reduced to a single integral of the general form
\indentdisplays{150pt}  

\begin{equation} 
E_{xc}[n]=\int d^3r n \epsilon_{xc}(n,\nabla n, \tau),
\end{equation}
where $n=\sum_{i,\sigma}^{occ.} |\Psi_{i,\sigma}|^2$ is  the electron density, $\nabla n$ its gradient, $\tau_{\sigma}=\sum_{i}^{occ.} |\Psi_{i,\sigma}|^2$ the positive orbital kinetic energy density, and $\Psi_{i,\sigma}$ are the Kohn-Sham orbitals.
Nonempirical functionals are generally built to satisfy exact constraints as far as possible.
It is here that the SCAN meta-GGA  \cite{PhysRevLett.115.036402} makes a substantial advance as it is the only semilocal exchange-correlation functional which satisfies the complete set of 17 known exact constraints that can be satisfied by semilocal functionals.
Moreover, SCAN is 'appropriately normed' in that it accurately captures interactions in rare-gas atoms and unbonded systems (see Supplementary Material of Sun \emph{et al.} \cite{PhysRevLett.115.036402}).
The earlier nonempirical meta-GGAs such as the Tao-Perdew-Staroverov-Scuseria (TPSS) \cite{tao2003climbing} and revTPSS \cite{PhysRevLett.103.026403} meta-GGA have been shown to be less accurate than the Perdew-Burke-Ernzerhof (PBE) GGA for the critical pressures of structural phase transitions of solids \cite{PhysRevB.74.121102, PhysRevB.88.184103}. SCAN meta-GGA eliminates this problem by introducing the dimensionless parameter

\begin{equation} 
\alpha=(\tau-\tau_{W})/\tau_{unif}>0,
\end{equation}
where $\tau_{W}=|\nabla n|^2/8n$ is the single-orbital limit of $\tau$, and $\tau_{unif}=(3/10)(3\pi^2)^{2/3}n^{5/3}$ is the uniform density limit.
The case of $\alpha=0$ corresponds to covalent single bonds while $\alpha\approx1$ to metallic, and the $\alpha\gg1$ limit describes weak bonds.
The rare-gas-atom norm contains information about $0<\alpha < \infty$, and some information about $\alpha\gg1$, while the non-bonded-interaction norm (the compressed Ar${_2}$) provides more information about $\alpha\gg1$.

Recently, SCAN meta-GGA has been tested in diversely bonded systems \cite{sun2016accurate}, where it was shown to be sophisticated enough to model a wide range of physical structures without being fitted to any bonded system.
In the present work, we apply it further to the class of thin film materials and we show a similar trend of successful predictions of ground-state structural and electronic properties. In particular, SCAN improves the overall agreement with experiment compared to LDA and GGA, at a comparable computational cost.
\

\section*{Computational Details} 
\

We have performed first-principles calculations using the pseudopotential projector augmented-wave method \cite{PhysRevB.59.1758} as implemented in the Vienna Ab-Initio Simulation Package (VASP) \cite{PhysRevB.54.11169,PhysRevB.48.13115}, with a kinetic energy cutoff of 400 eV (TMD monolayers and Bi$_2$Se$_3$ quintuple layer) and 800 eV (graphene, silicene, germanene, and phosphorene) for the plane-wave basis set.
The exchange-correlation functional was treated using LDA \cite{PhysRevB.23.5048, PhysRevLett.115.036402}, GGA-PBE \cite{PhysRevLett.77.3865,barbiellini1990effects} and SCAN meta-GGA \cite{PhysRevLett.115.036402}.
A 12 $\times$12$\times$1 $\Gamma$-centered $k$-point mesh was used to sample the Brillouin zone.
Spin-orbit coupling effects were included in the case of TMD monolayers and Bi$_2$Se$_3$ quintuple layer in a self-consistent manner.
We used a vacuum layer of at least 15 {\AA} thickness in the z-direction to simulate the films.
The equilibrium positions of the ions were calculated via structural optimization, where the internal degrees of freedom, along with the shape and volume of the unit cell, were allowed to vary until the residual forces per atom were less than 0.005 eV/\AA.
The resulting equilibrium unit cell was subsequently expanded and compressed uniformly around the equilibrium volume, while keeping the shape of the unit cell fixed.  
The equilibrium lattice constants were calculated by fitting the total energy per cell as a function of volume using the Birch-Murnaghan \cite{PMC1078704,PhysRev.71.809} equation of state:

\indentdisplays{130pt}  

\begin{equation} E\left ( V \right )=E_{0}+\frac{B_0 V_0}{B_{0}^{'}}\left [ B_{0}^{'}\left ( 1-\frac{V}{V_0} \right )+\left ( \frac{V_0}{V} \right )^{B_{0}^{'}-1} \right ],  \end{equation}

\noindent where $E$ is the total energy per cell, $E_0$ the equilibrium total energy per cell, $B_0$ the equilibrium bulk modulus, $V$ the unit cell volume, $V_0$ the equilibrium unit cell volume and $B_0^{'}$ the first derivative  of the bulk modulus with respect to $V$.
In this way, we determine $V_0$ (from which the equilibrium lattice constant $a$ was extracted), $B_0$ and $B_0^{'}$.
It should be noted that we are extending the Murnaghan fit to 2D materials, and quantities such as the bulk modulus should be regarded as fitting parameters rather than  physical quantities as discussed by Behera and Mokhopadhyay [BM]\cite{paper}. BM simulated the 2D-hexagonal structure of graphene and silicene using 3D-hexagonal supercells with large values of  the lattice parameter c to keep the interlayer interaction negligibly small.  They calculated for fixed values of $a$ the values of $c$ and the ground state energy $E_0$ for various cell volumes $V$, corresponding to different in-plane lattice constants $a$. Then, by fitting $E_0$ as a function of $V$ with the Birch-Murnaghan equation of state, they extracted $a$ from the value of $V$ at the minimum of $E_0$. Finally, the value of the lattice constant $a$ corresponding to $c$ going to infinity, was obtained by a linear fit of the data set ($a$, $1/c$).
Here, we have followed a similar procedure.

\

\section*{Results and Discussion}
\

\begin{figure}[H]
\centering
\includegraphics[width=\linewidth]{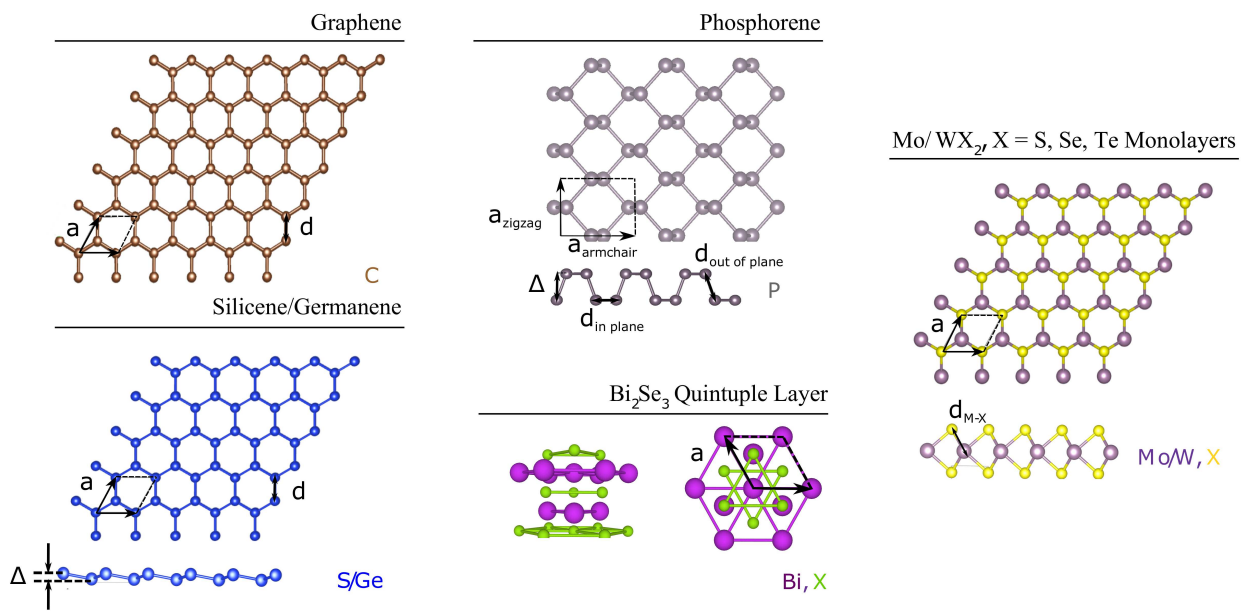}
\caption{Crystal structure of graphene in top view, and of silicene, germanene, phosphorene, {Bi}$_2${Se}$_3$ quintuple layer and {MX}$_2$ monolayers in top and side views.}
\label{fig:Fig1}
\end{figure}

\begin{figure}[H]
\centering
\includegraphics[width=1\linewidth]{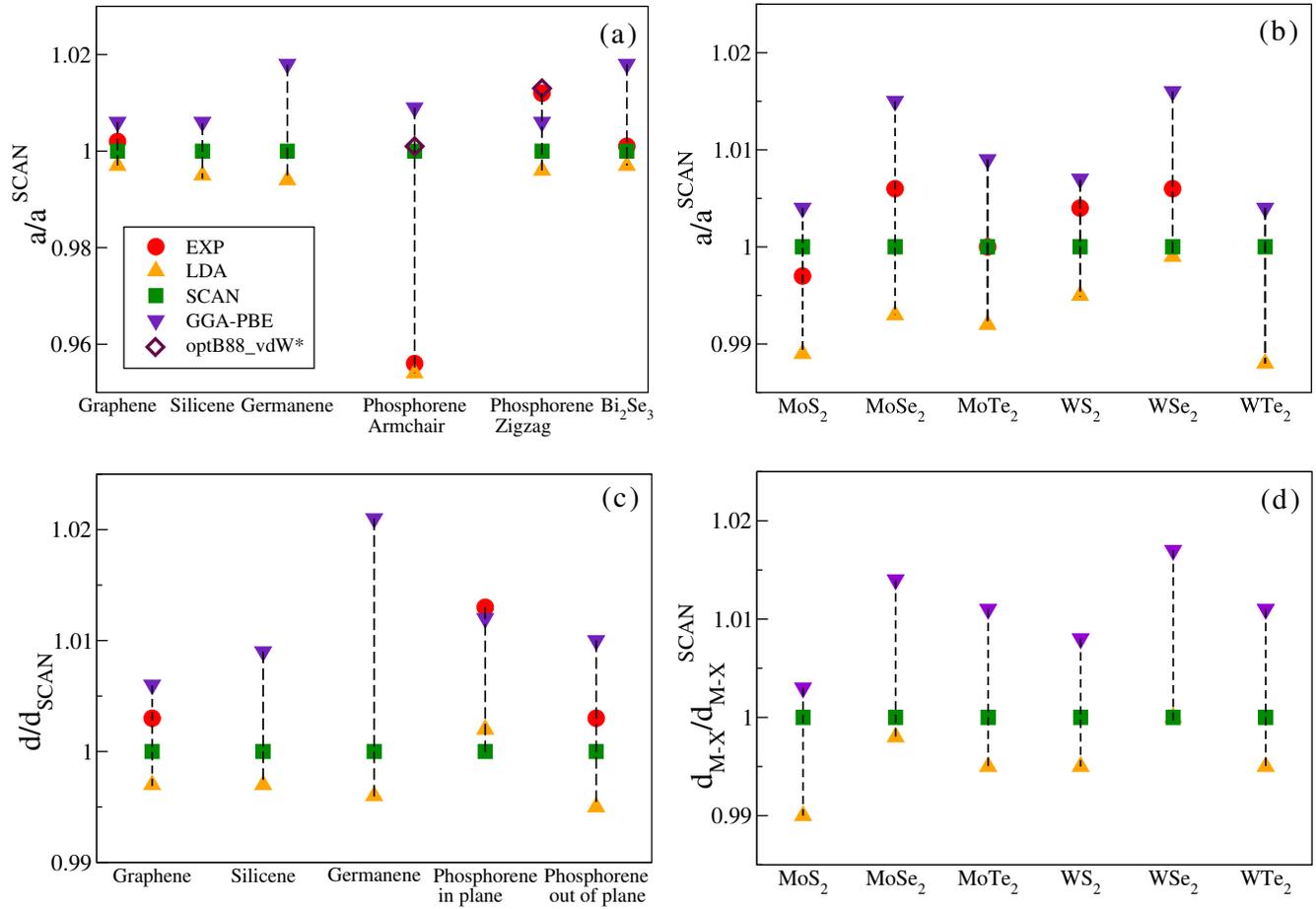}
\caption{Comparison between the calculated structural parameters using different exchange-correlation functionals: (a) and (b) Lattice constants, $a$. (c) and (d) Nearest-atom bond lengths, $d$, and transition metal-chalcogen distances, $d_{M-X}$. The experimental values for phosphorene are for the single-crystal black phosphorus compound. *The values of optB88$\char`_$vdW were taken from the work of Qiao \emph{et al.} \cite{qiao2014high}}.
\label{fig:Fig2}
\end{figure}

\begin{figure}[H]
\centering
\includegraphics[width=1\linewidth]{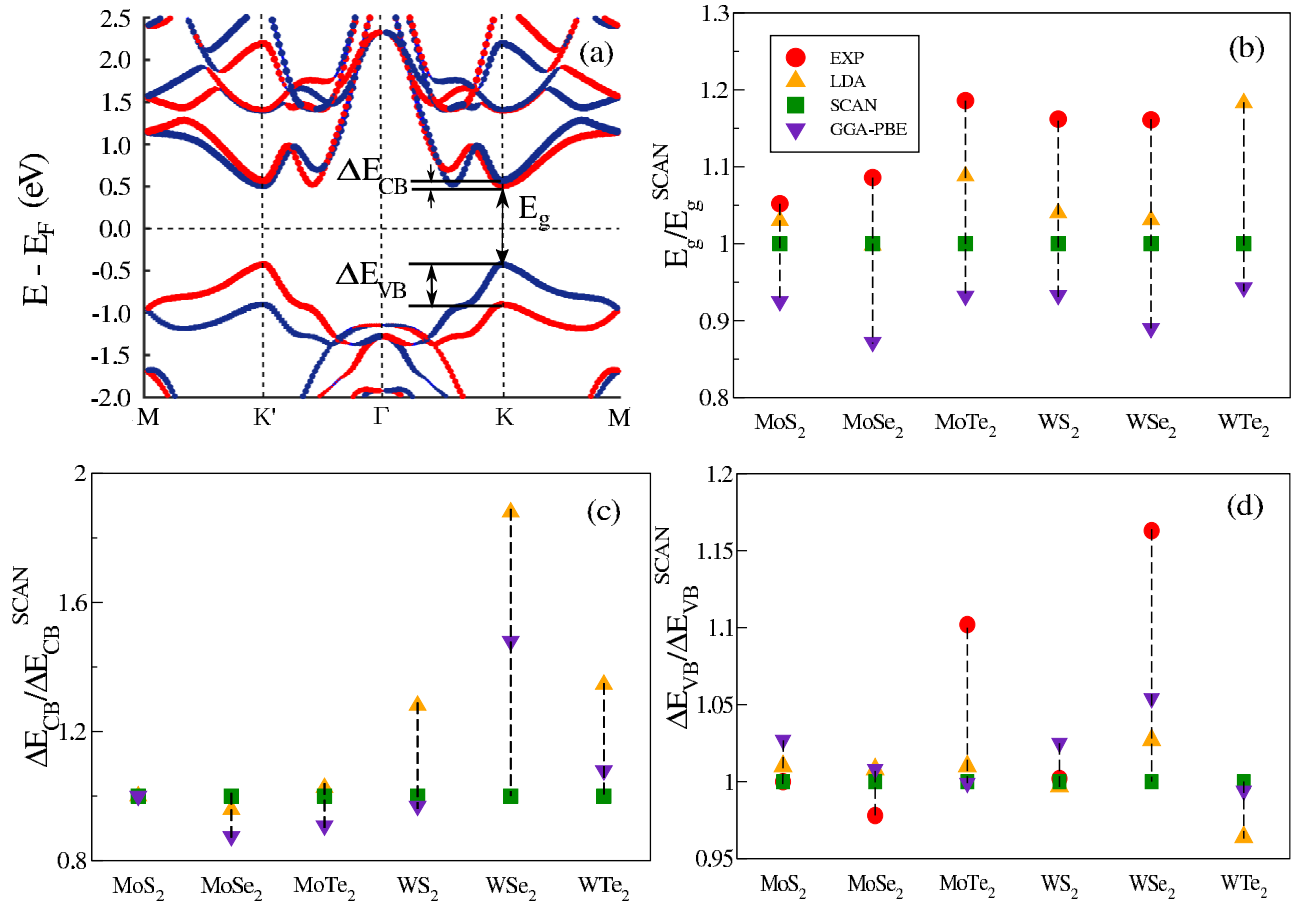}
\caption{(a) {WTe}$_2$ monolayer band structure along the high-symmetry lines  $M$-$K'$-$\Gamma$-$K$-$M$ in the Brillouin zone. The colored dots denote spin polarization: blue is for spin \emph{up}, and red is for spin \emph{down}. (b) Energy band gap values for the {MX}$_2$ monolayers, calculated within the LDA, GGA, and SCAN. The experimental values are for optical band gaps obtained from photoluminescence experiments. (c) and (d) Spin-splittings of the conduction and valence bands at the $K$ point, respectively. References for the experimental values in this figure are given in Table \ref{tab:example3}.}
\label{fig:Fig3}
\end{figure}

We present ground-state structural and electronic properties of a series of free-standing monolayer (ML) materials, which are: graphene, silicene, germanene and phosphorene, TMD monolayers {MX}$_2$ in the semiconducting 2H phase \cite{kappera2014phase} ({M} = {Mo}, {W}; {X} = {S}, {Se}, {Te}), and one quintuple layer (QL) film of {Bi}$_2${Se}$_3$. The crystal structures are depicted in Figure \ref{fig:Fig1}.
We tested how SCAN performs compared to the LDA and PBE-GGA by calculating the lattice constants $a$, the nearest-atom bond lengths $d$ for graphene, silicene, germanene and phosphorene, the buckling heights $\Delta$ for silicene, germanene and phosphorene, and X-M distances $d_{M-X}$ for the TMD monolayers (Figure \ref{fig:Fig2}).
These parameters are defined in Figure \ref{fig:Fig1}, and their values are given in Tables \ref{tab:example1} and \ref{tab:example2}.
For the TMD monolayers, we also extracted the band gaps $E_g$, as well as the spin-splittings at the $K$ point of the conduction band $\Delta E_{CB}$, and the valence band $\Delta E_{VB}$, as defined in Figure \ref{fig:Fig3}.

Table \ref{tab:example1} lists values of lattice constants $a$, bulk moduli  $B_0$ and their first derivatives $B'_0$, the last two being fitting parameters as we discussed in the Computational Details section above.
Trends in the lattice constants are visualized in frames (a) and (b) of Figure \ref{fig:Fig2}. The LDA is seen to underestimate $a$, in agreement with the expectation that it leads to overbinding in solids \cite{RevModPhys.61.689}.
On the other hand, the GGA overcorrects $a$, especially for heavier elements as seen by comparing germanene with silicene and graphene in Figure \ref{fig:Fig1}(a), a behavior observed in 3D metals more generally \cite{barbiellini1990effects}.
Figs. \ref{fig:Fig2}(a) and 2(b) show that the SCAN meta-GGA values lie between the LDA and GGA predictions, suggesting that SCAN meta-GGA cures the overcorrection of the GGA, and generally yields better agreement with experiment, within about 0.5\%, although experimental data on freestanding silicene, germanene and phosphorene are not currently available. Remarkably, for the QL {Bi}$_2${Se}$_3$, the SCAN-based lattice parameter is also in excellent accord with the experimental value reported by Kou \emph{et al.} \cite{doi:10.1021/nl4037214}.

The role of electron correlations in graphene remains an open problem. Accurate Quantum Monte Carlo (QMC) simulations suggest that the ground state of graphene is highly nontrivial, with significant contributions from resonating valence bond (RVB) type states \cite{PhysRevLett.107.086807}. [RVB effects appear to be important in systems of low dimensionality more generally, such as the Li clusters \cite{PhysRevB.79.035416}.] The fact that SCAN reproduces the experimental lattice constant of graphene quite well thus indicates that SCAN can reasonably capture features of complex ground states in 2D systems.

It is interesting to consider the QMC result for the lattice constant along the armchair direction in phosphorene \cite{doi:10.1021/acs.nanolett.5b03615}. Surprisingly, we find that LDA already overestimates the QMC armchair lattice constant, even though one normally expects overbinding from the LDA. [Note, experimental lattice constants for phosphorene are not currently available.] Furthermore, we find that SCAN also overestimates the QMC result, as seen in Table \ref{tab:example1}, and performs at the level of the optB88-vdW \cite{qiao2014high} functional, see Fig. \ref{fig:Fig2}(a).
It is not clear to what extent relaxing the fixed-node approximation in QMC might expand the armchair lattice constant in monolayer black phosphorus, and restore the usual paradigm of LDA underestimating lattice constants more generally.
We emphasize that when phosphorene layers are coupled, it becomes crucial to include van der Waals corrections. For example, in bulk black phosphorus, SCAN+rvv10 yields lattice constants in close agreement with both experiment and QMC \cite{PhysRevX.6.041005}, and represents a considerable improvement over PBE+vdW.
Concerning the phosphorene lattice constant along the zigzag direction, our results in Table \ref{tab:example1} show that  is fairly insensitive to corrections beyond the LDA.


Table \ref{tab:example2} shows that the equilibrium structures assumed by all 2D films considered (other than graphene) are buckled, i.e. exhibit non-zero values of $\Delta$, and that the buckling is amplified in going from the LDA to the GGA.
In sharp contrast, SCAN predicts smaller buckling heights for silicene and germanene compared to the LDA.
A possible reason for this flattening trend is that SCAN satisfies the non-uniform coordinate scaling constraint \cite{PhysRevLett.115.036402}, while the LDA and GGA do not.
In phosphorene, since the buckling height is much larger than that in silicene and germanene, and lies at the scale of a typical chemical bond, SCAN predicts a value comparable to LDA and GGA.
Turning to bond lengths, here also we see that, like the lattice constants, SCAN systematically rectifies GGA's tendency to overcorrect LDA, see Table \ref{tab:example2} and Figs. \ref{fig:Fig2}(c) and \ref{fig:Fig2}(d).
For the TMD films trends in bond lengths between the LDA, SCAN and GGA are similar. Notably, spin-orbit effects, which are included in the calculations, do not seem to influence the trends in bond lengths.

Given the interest in potential applications of TMD films \cite{doi:10.1021/jz502646d}, Fig. \ref{fig:Fig3}(a) shows the band structure of a {WTe}$_2$ monolayer, which is typical of the family of TMD monolayers considered.
Table \ref{tab:example3} gives the band gaps obtained from the band structures based on different functionals computed \emph{at the equilibrium crystal structures}, see also Fig. \ref{fig:Fig3}(b).
Note that our band structures arise from a ground-state theory \cite{BandGaps, RevModPhys.61.689}, and thus do not accurately model the band gaps.
Nevertheless, the LDA is well-known to reasonably capture optical energy gaps in many materials.
GGA expands the lattice, and it generally worsens the band gap.
In contrast, consistent with the findings of Yang \emph{et al.} \cite{PhysRevB.93.205205}, SCAN restores an improved agreement with the experimental band gaps, together with improved lattice structures.
This good agreement can be understood to be a result of using the generalized Kohn-Sham theory \cite{BandGaps} within SCAN meta-GGA.
Incidentally, within the many-body body perturbation theory, Qiu ${et. al}$ \cite{PhysRevLett.115.176801} have noticed an interesting compensation between the quasiparticle (QP) and excitonic corrections in the case of transition metal dichalcogenides.
For example, in {MoS}$_2$, the GW approximation yields a direct gap of 2.67 eV.
The observed optical gap is about 0.8 eV smaller, which could be explained as the exciton binding energy.

Returning to Figure \ref{fig:Fig3}(a), note that the conduction band (CB) and the valence band (VB) are split at the $K$-point, which is a consequence of spin-orbit coupling \cite{Chang2014, Alidoust2014}.
Furthermore, because TMD monolayers lack inversion symmetry, there is an inversion in the spin-resolved band structures near the Fermi level between the $K$ and $K'$ symmetry points (Fig. \ref{fig:Fig3}(a)), where blue dots denote spin \emph{up} and red dots spin \emph{down}.
These features of band structures of TMD monolayers have been predicted in earlier DFT calculations \cite{PhysRevB.88.245436} and observed in  experiments \cite{kormanyos2015k, Zhang2014, PhysRevB.64.235305, PhysRevB.92.245442, 0953-8984-27-18-182201}.
We define the CB and VB spin-splitting energies as: $\Delta E_{CB}=E_{CB}^{\uparrow}-E_{CB}^{\downarrow}$, and $\Delta E_{VB}=E_{VB}^{\uparrow}-E_{VB}^{\downarrow}$.
The values of these splitting energies are listed in Table \ref{tab:example3}.
We see that $\Delta E_{CB}<0$ for {MoX}$_2$ monolayers, and $\Delta E_{CB}>0$ for the {WX}$_2$ counterparts.
This sign change can be explained in terms of the material-dependent spin-orbit coupling effects \cite{PhysRevB.88.245436}.
The results of Table \ref{tab:example3} indicate that SCAN predicts the spin-splittings in TMD monolayers, at least in some cases ({MoS}$_2$, {MoSe}$_2$ and {WS}$_2$), more accurately than the LDA and GGA (see Figure \ref{fig:Fig3}(d)).
We thus adduce that SCAN reasonably describes the delicate balance between the exchange, correlation and spin-orbit coupling interactions, which underlie spin-resolved band structures. The exquisite ability of SCAN to capture such subtle effects will allow the study of controlled magnetism in 2D crystals. Interesting proposals have been put forward for monolayer transition metal dichalcogenides \cite{PhysRevB.87.100401}, but magnetic order has not been proven so far in experiments. SCAN meta-GGA could thus accelerate the discovery of these fascinating materials.

\section*{Conclusions}

In order to test the efficacy of the recently proposed SCAN functional toward capturing improved ground state properties of layered materials, we have carried out SCAN based computations on monolayers of graphene and a number of 'beyond graphene' compounds, including films of transition-metal dichalcogenides (TMDs).
The results are compared and contrasted with those based on the commonly used LDA and GGA schemes.
SCAN is shown to yield systematic improvements in the equilibrium lattice constants and the nearest-atom bond lengths.
We also consider band gaps and spin-splittings in the TMD films, and show that here also the SCAN functional leads to improvements, difficulties of interpreting band gaps in a ground state computation notwithstanding.
We thus conclude that SCAN would provide an improved description of the ground-state electronic and geometric structures of layered materials more generally, at a cost comparable to the LDA and GGA. 

\bibliography{References.bib}

\section*{Acknowledgements}

It is a pleasure to thank Professor J. P. Perdew  for important discussions.
This work was supported by the US Department of Energy (DOE), Office of Science, Basic Energy Sciences grant number DE-FG02-07ER46352 (core research), and benefited from Northeastern University's Advanced Scientific Computation Center (ASCC), the NERSC supercomputing center through DOE grant number DE-AC02-05CH11231 and  the support (applications to layered materials) from the DOE EFRC: Center for the Computational Design of Functional Layered Materials (CCDM) under grant number DE-SC0012575.

\section*{Author contributions statement}
B.B., J.S. and C.L. conceived and initiated the study. J.S. and A.R. provided the SCAN meta-GGA functional. I.G.B. performed the calculations and analysis. I.G.B., C.L., B.B. and A.B. prepared the manuscript. All authors  contributed to the discussions and reviewed the manuscript.
\section*{Additional information}
The authors declare no competing financial interests.

\begin{table}[ht]
\centering
\begin{tabular}{lllllll}
\hline
\\

 & Graphene & Silicene & Germanene & $\ \ \ \ \ \ \ $Phosphorene\ \ \ \ \ \ & QL\ {Bi}$_2${Se}$_3$ \\
  &   &   &   &armchair \ \ \ \ \ \ \ \ zigzag & \\
\hline
\hline
\\ 

$a^{EXP}(\AA)$ & 2.459 \cite{PhysRev.100.544} & - & - &\ \ 4.322 \cite{doi:10.1021/acs.nanolett.5b03615}$\ \ \ \ \ \ $*3.314 \cite{Brown:a04860} &	4.138 \cite{doi:10.1021/nl4037214}\\

$a^{LDA}(\AA)$ & 2.445 & 3.860 & 3.981 & \ \ 4.365 $\ \ \ \ \ \ \ \ \ $ 3.264 & 4.122\\

$a^{SCAN}(\AA)$ & 2.453 & 3.879 & 4.004 & \ \ 4.576 $\ \ \ \ \ \ \ \ \ $ 3.276 & 4.135 \\

$a^{PBE}(\AA)$ & 2.468 & 3.902 & 4.077 & \ \ 4.618 $\ \ \ \ \ \ \ \ \ $ 3.297 & 4.210\\
\\

\hline
\\

$B_0^{LDA}(GPa)$ & 34.139 &3.922 & 20.717 & \ \ \ \ \ \ \ \ \ \ 34.458 &22.302\\

$B_0^{SCAN}(GPa)$ & 33.914 &4.049 & 21.142 &  \ \ \ \ \ \ \ \ \ \ 34.026 & 21.166\\

$B_0^{PBE}(GPa)$ & 32.538 &3.651 & 15.418 &  \ \ \ \ \ \ \ \ \ \ 30.600 & 18.694 \\
\\

\hline
\\

$B_0'^{LDA}$ & 3.363&4.505 & 6.061&  \ \ \ \ \ \ \ \ \ \ 5.105 & 4.564\\

$B_0'^{SCAN}$ & 3.348&4.243 &9.057 &  \ \ \ \ \ \ \ \ \ \ 5.370 & 4.743\\

$B_0'^{PBE}$ & 3.323&4.304 &6.514 &  \ \ \ \ \ \ \ \ \ \ 5.662 &  4.456\\
\\

\hline
\hline
*bulk value\\

\\

\hline
\\

 & ML {MoS}$_2$ & ML {MoSe}$_2$ & ML {MoTe}$_2$ & \ \ \ \ \ \ \ \ \ \ ML {WS}$_2$ &ML {WSe}$_2$ & ML {WTe}$_2$ \\
\\
 
\hline
\hline
\\ 

$a^{EXP}(\AA)$ & 3.160 \cite{kormanyos2015k} & 3.288 \cite{kormanyos2015k} & 3.519 \cite{kormanyos2015k}& \ \ \ \ \ \ \ \ \ \ 3.154 \cite{kormanyos2015k}& 3.286 \cite{kormanyos2015k}& 3.496 \cite{iwantocitethis} \\
$a^{LDA}(\AA)$ & 3.136&3.245&3.490&\ \ \ \ \ \ \ \ \ \ 3.127&3.263&3.498\\

$a^{SCAN}(\AA)$&3.170&3.268&3.518&\ \ \ \ \ \ \ \ \ \ 3.142&3.265&3.542\\

$a^{PBE}(\AA)$ &3.181&3.318&3.551&\ \ \ \ \ \ \ \ \ \ 3.165&3.317&3.557\\
\\

\hline
\\

$B_0^{LDA}(GPa)$ & 56.992&356.800&37.696&\ \ \ \ \ \ \ \ \ \ 61.712&373.280&40.128
\\

$B_0^{SCAN}(GPa)$&54.080&358.400&37.248&\ \ \ \ \ \ \ \ \ \ 60.576&375.520&37.840
\\

$B_0^{PBE}(GPa)$ &50.048&305.600&32.336&\ \ \ \ \ \ \ \ \ \ 55.392&332.960&34.944\\
\\

\hline
\\

$B_0'^{LDA}$ & 4.341 & 5.742 & 4.573 &\ \ \ \ \ \ \ \ \ \ 4.486 &5.799
 & 4.512 \\

$B_0'^{SCAN}$ & 5.293 & 5.307 & 4.383 &\ \ \ \ \ \ \ \ \ \ 5.252 & 11.299
 & 7.263 \\

$B_0'^{PBE}$ & 4.533 & 5.964& 4.507 &\ \ \ \ \ \ \ \ \ \ 5.067 & 5.669
 & 4.471\\
\\

\hline
\hline

\end{tabular}
\caption{\label{tab:example1} Ground-state lattice constants $a$, bulk moduli $B_0$ and first derivatives $B'_0$, calculated by fitting the total energy per cell with the Birch-Murnaghan equation of state. Notably, there are two sources of error in the determination of the lattice constants, both of which are of the order of $0.0005 \AA$. One originates from the Murnaghan fit, taken as $\Delta a=|a^M-a^B|$, where $a^M$ is the Murnaghan fit lattice constant, and $a^B$ is the Birch lattice constant obtained by applying the constraint $B'_0$ = 4. The second source of error is the interpolation of the vacuum layer c to $\infty$, as calculated by Behera and Mokhopadhyay \cite{paper}.}
\end{table}

\begin{table}[ht]
\centering
\begin{tabular}{lllllll}
\hline
\\

 & &Graphene & Silicene & Germanene & $\ \ \ \ \ \ \  $Phosphorene \\
\\

\hline
\hline
\\ 
 & & & & & in plane \ \ \ \ \ out of plane \\
&$d^{EXP}(\AA)$ &1.420 \cite{RevModPhys.81.109}  & - & - &*2.224 \cite{Brown:a04860}$\ \ \ \ $ \ *2.244 \cite{Brown:a04860} \\

&$d^{LDA}(\AA)$ &1.412  &2.250  & 2.380 & \ \ 2.199 $\ \ \ \ \ \ \ \ \ $ 2.225 \\

&$d^{SCAN}(\AA)$ &1.416  &2.256  &2.391 &\ \  2.195 $\ \ \ \ \ \ \ \ \ $ 2.237  \\

&$d^{PBE}(\AA)$ &1.424 &2.277  &2.442&\ \  2.220 $\ \ \ \ \ \ \ \ \ $ 2.259 \\
\\

\hline
\\

&$\Delta^{LDA}(\AA)$ &  &0.437  & 0.647 & \ \ \ \ \ \ \ \ \ \ 1.942 \\

&$\Delta^{SCAN}(\AA)$ & &0.383  &0.634 & \ \ \ \ \ \ \ \ \ \ 2.070\\

&$\Delta^{PBE}(\AA)$ & &0.442  &0.660& \ \ \ \ \ \ \ \ \ \ 2.070\\
\\

\hline
\hline
*bulk values\\
\\ 

\hline
\\

 & ML {MoS}$_2$ & ML {MoSe}$_2$ & ML {MoTe}$_2$ &  ML {WS}$_2$ & \ \ \ \ \ \ \ ML {WSe}$_2$ & ML {WTe}$_2$ \\
\\

\hline
\hline
\\

$d_{M-X}^{LDA}(\AA)$ & 2.381 & 2.502 & 2.689 & 2.385&\ \ \ \ \ \ \ \ 2.503 &2.695\\

$d_{M-X}^{SCAN}(\AA)$ &2.405  & 2.506 &2.703 &2.397    & \ \ \ \ \ \ \ \ 2.504&2.705\\

$d_{M-X}^{PBE}(\AA)$ &2.413 &2.541 &2.733  &2.416   &\ \ \ \ \ \ \ \ 2.546&2.738 \\
\\

\hline
\\
\end{tabular}
\caption{\label{tab:example2}Nearest-atom bond lengths $d$ and buckling heights $\Delta$ for graphene and 'beyond graphene' materials, as well as ${M-X}$ bond lengths $d_{M-X}$ for the TMD monolayers.}
\end{table}

\begin{table}[ht]
\centering
\begin{tabular}{lllllll}

\hline
\\

 & ML {MoS}$_2$ & ML {MoSe}$_2$ & ML {MoTe}$_2$ & ML {WS}$_2$ &ML {WSe}$_2$ & ML {WTe}$_2$ \\
\\
 
\hline
\hline
\\

$E_g^{EXP}(eV)$ & 1.830 \cite{C4CS00301B}&1.660  \cite{C4CS00301B}&1.100 \cite{PhysRevB.94.085429}&1.950  \cite{C4CS00301B}&1.640 \cite{C4CS00301B}&-\\

$E_g^{LDA}(eV)$ & 1.792  & 1.526  & 1.102  & 1.745  & 1.456  & 0.933 \\

$E_g^{SCAN}(eV)$&1.740&1.529&1.013&1.678&1.412&0.788\\

$E_g^{PBE}(eV)$ &1.590&1.340&0.947&1.580&1.270&0.765\\
\\

\hline
\\
$\Delta E_{VB}^{EXP}(meV)$ & 145$\pm$4 \cite{PhysRevLett.114.046802}&180\cite{Zhang2014}&238$\pm$10 \cite{PhysRevB.64.235305}&419$\pm$11 \cite{PhysRevB.92.245442} & 513$\pm$10 \cite{0953-8984-27-18-182201} &-\\

$\Delta E_{VB}^{LDA}(meV)$ & 147 & 186  & 218  & 417  & 453  & 472 \\

$\Delta E_{VB}^{SCAN}(meV)$&145&184&216&418&441&489\\

$\Delta E_{VB}^{PBE}(meV)$ &148&186&213&425&462&480\\
\\

\hline
\\

$\Delta E_{CB}^{LDA}(meV)$ & -3  & -23  & -39  & 41  & 47  & 65 \\

$\Delta E_{CB}^{SCAN}(meV)$&-3&-24&-38&32&25&48\\

$\Delta E_{CB}^{PBE}(meV)$ &-3&-20&-32&31&37&54\\
\\

\hline
\end{tabular}
\caption{\label{tab:example3}Values of the energy band gaps $E_g$, along with the corresponding optical gaps obtained from photoluminescence experiments $E^{EXP}_g$. Also given are the spin-splitting energies at the $K$ point in the valence band $\Delta E_{VB}$, and the conduction band $\Delta E_{CB}$ for the TMD monolayers.}
\end{table}

\end{document}